\documentclass{rspublic}
\usepackage{epsfig}
\newcommand{\be}{\begin{equation}}
\newcommand{\ee}{\end{equation}}
\newcommand{\etal}{{\em et al.}}
\newcommand{\ltorder}{\hbox{ \rlap{\raise 0.425ex\hbox{$<$}}\lower
0.65ex\hbox{$\sim$} }}
\newcommand{\gtorder}{\hbox{ \rlap{\raise 0.425ex\hbox{$>$}}\lower
0.65ex\hbox{$\sim$} }}

\newcommand{\apj}{{\it Astrophys. J.}}
\newcommand{\apjs}{{\it Astrophys. J. Suppl. Ser.}}
\newcommand{\aj}{{\it Astron. J.}}
\newcommand{\mnras}{{\it Mon. Not. R. Astron. Soc.}}
\newcommand{\aanda}{{\it Astron. Astrophys.}}
\newcommand{\ptra}{{\it Phil. Trans. R. Soc. A}}

\begin{document}

\title[Star cluster dynamics]{Star cluster dynamics}

\author[E. Vesperini]{Enrico Vesperini}

\affiliation{Department of Physics, Drexel University, Philadelphia, PA 19104, USA}

\label{firstpage}

\maketitle

\begin{abstract}{globular clusters: general, $N$-body simulations, stellar dynamics}
Dynamical evolution plays a key role in shaping the current properties
of star clusters and star cluster systems. A detailed understanding of
the effects of evolutionary processes is essential to be able to
disentangle the properties which result from dynamical evolution from
those imprinted at the time of cluster formation.  In this review, we
focus our attention on globular clusters and review the main physical
ingredients driving their early and long-term evolution, describe the
possible evolutionary routes and show how cluster structure and
stellar content are affected by dynamical evolution.
\end{abstract}

\section{Introduction}\label{sec:intro}

Globular star clusters have long been considered the ideal
astrophysical objects to explore many aspects of stellar dynamics and,
in particular, to study the evolution of stellar systems as driven by
the effects of two-body relaxation.

Only recently, however, the actual complexity of globular cluster
dynamics has emerged from the wealth of new observational and
theoretical studies that have clearly shown the close interplay
between stellar dynamics, stellar evolution, the clusters' stellar
content and the dynamics and properties of the host galaxy.  The
notion of an `ecology of star clusters', introduced by Heggie (1992),
nicely illustrates the close interplay among the different elements of
the astrophysics of star clusters.

To use star clusters as tools to advance and guide our understanding
of star formation in galaxies, it is essential to understand to what
extent the current properties of star clusters and star cluster
systems are the result of evolutionary processes. We also need to be
able to discern the signatures of properties imprinted by formation
processes from those determined by dynamical evolution. In this paper,
we will review the main physical ingredients driving the early and
long-term evolution of clusters and describe how cluster structure and
stellar content is affected by dynamical evolution.

In \S\ref{sec:early}, we discuss the early evolution of clusters,
i.e., the processes that can affect their stellar content early in
their life and lead to their rapid dissolution.  In
\S\ref{sec:longterm}, we review the long-term evolution of clusters as
driven, primarily, by two-body relaxation, focusing our attention on
the evolution of cluster structure, mass and stellar
content. Conclusions are summarized in \S\ref{sec:concl}.

\section{Early dynamical evolution}\label{sec:early}

Although much progress has been made in recent years (see, e.g.,
Clarke 2010; Lada 2010), a full understanding of the formation and the
initial structural properties of star clusters is still lacking.

Analytical and numerical studies of globular cluster evolution often
adopt initial conditions which resemble the current structural
properties of clusters (e.g., purely stellar, spherical and isotropic
King models), rather than the possibly more complex structures
suggested by theoretical and observational studies of very young star
clusters (e.g., Elmegreen 2000; Bonnell \etal\ 2003; Allen \etal\
2007).  Figure 1 shows a few snapshots of a hydrodynamical simulation
modelling the formation of a small star cluster (Bonnell \etal\ 2003)
and nicely illustrates the clumpy and irregular structure of a young
star cluster still embedded in the primordial gas from which its stars
are forming.

In addition, the standard paradigm according to which clusters are
`simple stellar populations' composed of stars of the same age and
chemical composition has recently been challenged by the increasing
observational evidence of the presence of multiple stellar populations
in globular clusters (e.g., Bedin \etal\ 2004; Gratton \etal\ 2004;
Piotto \etal\ 2007; Carretta \etal\ 2008, 2009{\it a,b}; D'Antona \&
Caloi 2008; Milone \etal\ 2008; see also the reviews of van Loon 2010;
Kalirai \& Richer 2010; and Bruzual 2010). This is a major paradigm
shift and significantly increases the complexity of models following
the formation and dynamical evolution of globular clusters (e.g.,
D'Ercole \etal\ 2008; Renzini 2008; and references therein).

Although the exact structural and kinematic properties of newly formed
clusters are still uncertain, a number of studies have explored the
early evolution of clusters and shown their possible evolutionary
routes.  In the following subsections, we present an overview of the
early dynamical evolution of clusters, based on analytical
calculations and $N$-body simulations.

\begin{figure}[h]
\centering{
\includegraphics[width=7cm]{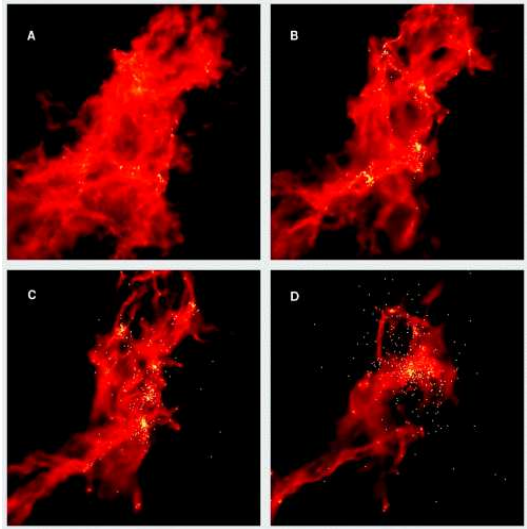}
}
\caption{Formation and early evolution of a star cluster from a
  hydrodynamical simulation following the hierarchical fragmentation
  of a turbulent molecular cloud (from Bonnell \etal\ 2003).}
\label{fig1}
\end{figure}

\subsection{Evolution of clumpy systems}\label{sec:early_clumpy}

A number of observational and numerical studies suggest that star
clusters might form with irregular clumpy structures and thus out of
virial equilibrium. These studies mostly concern low-mass clusters and
it is not clear whether such initial conditions are shared by more
massive globular-cluster-like objects. Many clusters as young as $\sim
50-100$ Myr already exhibit smooth surface-brightness profiles (see,
e.g., Elson \etal\ 1987; Larsen 2004) which are well fit by the
`Elson, Fall and Freeman' profile (hereafter EFF; Elson \etal\ 1987),
\be
\Sigma \propto {1 \over (1+(r/r_\mathrm{c})^2)^{\gamma/2}},
\ee 
where $\Sigma$ is the cluster's surface brightness and $r_\mathrm{c}$
its core radius. A number of studies have investigated the dynamics of
clusters starting from clumpy initial conditions and explored if, and
on what timescale, early dynamical evolution can drive them towards
properties similar to those of older clusters with a smoother density
profile.

The evolution of clusters with different levels of clumpiness, clump
spatial distributions and initial velocity distributions has been
studied in general surveys (e.g., Aarseth \etal\ 1998; Goodwin 1998;
Goodwin \& Whitworth 2004; McMillan \etal\ 2007) as well as in
investigations aimed at modelling the evolution of specific clusters
(e.g., Scally \& Clarke 2002). Recently, Hurley \& Bekki (2008) made
further progress towards more realistic models and explored the
stellar dynamical evolution of an inhomogeneous, clumpy stellar system
with initial conditions produced by an hydrodynamical simulation of a
turbulent giant molecular cloud.

The results of all these numerical studies show that, in most cases,
an initially clumpy system undergoes a rapid phase of violent
relaxation, which erases any initial substructure and leaves the
cluster with a density profile which is very similar to the EFF
profile.  Similar simulations have also been carried out in the
context of early galaxy evolution (see, e.g., the first simulations by
van Albada 1982; McGlynn 1984; see Trenti \etal\ 2005, and references
therein, for a more recent study) and, in fact, even before the EFF
profile was used to fit the observed surface-brightness profiles of
young clusters in the Large Magellanic Cloud, it was adopted by
McGlynn (1984) in a study of early galaxy evolution to fit the density
profiles emerging at the end of the violent-relaxation phase of
initially clumpy systems.

The timescale to reach final equilibrium without any significant
remaining trace of the initial substructures depends on the details of
the initial conditions but it is, in general, $\approx
1-10~t_\mathrm{dyn}$ where $t_\mathrm{dyn}=\sqrt{3\pi/16G\rho}$ is the
cluster dynamical time (and $\rho$ the cluster's average density).
Fellhauer \etal\ (2009) recently showed that the clump-merging
timescale may be shorter than the timescale of primordial-gas
expulsion, thus enhancing the cluster's chances to survive the
disruptive effects of gas expulsion (see \S2\,$c$ below)

Statistical tools to more rigorously measure cluster substructure have
been used by Cartwright \& Whitworth (2004), Schmeja \& Klessen (2006)
and Kumar \& Schmeja (2007) and provided a quantitative measure of the
evolving level of clumpiness in observed young clusters and
star-forming regions, as well as in the numerical results of
hydrodynamical simulations of star cluster formation.

\subsection{Clumpy systems and initial mass segregation}\label{sec:early_segreg} 

Although the evolution of clumpy stellar systems is rapid and occurs
on timescales on the order of a few dynamical timescales, it can have
important implications for the generation of mass segregation early in
a cluster's evolution. Mass segregation, the tendency of more massive
stars to sink towards the cluster's central regions, is a consequence
of energy equipartition driven by two-body relaxation and one of the
processes normally associated with the long-term evolution of
clusters. However, a number of observational studies (e.g.,
Hillenbrand 1997; Hillenbrand \& Hartmann 1998; Fischer \etal\ 1998;
de Grijs \etal\ 2002; Sirianni \etal\ 2002; Gouliermis \etal\ 2004;
Stolte \etal\ 2006; Sabbi \etal\ 2008; Allison \etal\ 2009{\it a};
Gouliermis \etal\ 2009; see also Ascenso \etal\ 2009) have found
evidence of mass segregation in clusters with ages shorter than the
time needed to produce the observed segregation by two-body relaxation
(see also de Grijs 2010).

It has been suggested in a number of theoretical studies that the
observed segregation in young clusters would be primordial and
imprinted by the star-formation process (Bonnell \etal\ 1997, 2001;
Bonnell \& Davies 1998; Klessen 2001; Bonnell \& Bate
2006). Specifically, the mechanism responsible for primordial mass
segregation would be the higher accretion rate for stars in the
central regions of young clusters. The actual efficiency of this
mechanism is still a matter of debate (Krumholz \etal\ 2005; Krumholz
\& Bonnell 2007), but if the individual clumps are indeed mass
segregated, it has been shown by McMillan \etal\ (2007) that such
primordial mass segregation would not be erased in the
violent-relaxation phase during which clumps merge. The final system
would preserve the mass segregation of the original clumps (see also
Fellhauer \etal\ 2009; Moeckel \& Bonnell 2009).

In the same study, McMillan \etal\ also presented an alternative
scenario for a dynamical origin of early mass segregation in young
clusters. Even if the clumps are not initially segregated, if their
internal segregation timescale is shorter than the time needed for the
clumps to merge, they will segregate through standard two-body
relaxation and preserve this segregation after they have merged. The
multiscale dynamical evolution of clumpy systems is, in this case,
responsible for rapidly leading to mass segregation in young clusters
without invoking any mechanism associated with the star-formation
process.

In a recent study, Allison \etal\ (2009{\it b}) showed that mass
segregation can be rapidly produced dynamically also in the
high-density core formed during the collapse of cold clumpy clusters.
Bastian \etal\ (2008) found observational evidence of a strong
expansion in the first 20 Myr of evolution of six young M51 clusters
and pointed out that this expansion could also lead to a rapid
variation in the cluster relaxation time. As pointed out by Bastian
\etal, using the current relaxation time might lead to an
underestimate of the possible role played by two-body relaxation in
generating mass segregation in the early phases of a cluster's
dynamical evolution.

Regardless of the mechanism producing mass segregation, the presence
of segregation very early on in a cluster's life can have a
significant impact on its dynamical evolution. We will discuss some of
the implications of initial mass segregation in the next subsection.

\subsection{Early mass loss: primordial gas expulsion and stellar
evolution}\label{sec:early_gas} 

Expulsion of the primordial residual gas in which a young cluster is
still embedded in the very early stages of its life, along with mass
loss due to stellar evolution, can have a significant impact on a
cluster's structure and survival chances.

Early analytical calculations by Hills (1980) based on the virial
theorem showed that clusters in virial equilibrium losing impulsively
more than half of their mass would rapidly expand and dissolve in
response to this mass loss. Subsequent studies by Boily \& Kroupa
(2003{\it a,b}) refined Hills' results by calculating the fraction of
stars remaining bound in a cluster with a given velocity distribution
function after the impulsive removal of a given amount of gas. Boily
\& Kroupa's study showed that, for clusters with a high-binding-energy
massive core, up to 70\% of the cluster mass can be removed
impulsively without leading to complete cluster dissolution (see also
the simulations in Kroupa \etal\ 2001; and references therein).

Goodwin \& Bastian (2006) explored, by means of $N$-body simulations,
the evolution of the structural properties of clusters in the stages
following gas expulsion and showed that the observed luminosity
profiles of a few young massive clusters differ from EFF and King
models in their outer regions, consistent with their
simulations. These could, therefore, be examples of clusters currently
expanding in response to gas expulsion.

Baumgardt \& Kroupa (2007) carried out a survey of $N$-body
simulations to explore the dependence of the response to residual gas
expulsion on the star-formation efficiency, the expulsion timescale
and the strength of the host galaxy's tidal field. Slow gas removal
and a weak tidal field can increase the amount of mass a cluster can
lose without undergoing complete disruption. However, in most cases a
cluster undergoes significant expansion and needs to be initially much
more compact than observed today to survive this phase. As shown in a
number of numerical studies (e.g., Portegies Zwart \etal\ 2002, 2004;
G\"urkan \etal\ 2004; see also McMillan 2008, and references therein,
for a review), very high initial central densities might lead to a
rapid core collapse and segregation of massive stars, and trigger a
runaway merger of massive stars, leading to the formation of an
intermediate-mass black hole (IMBH; but see Glebbeek \etal\ 2009 for a
recent study of the stellar evolution of runaway merger products,
showing how mass loss due to stellar winds might significantly affect
this process and prevent the formation of an IMBH).

Early mass loss due to stellar evolution (for example, expulsion of
Type II supernova ejecta) can also have a significant impact on early
cluster evolution.  Semi-analytic calculations (e.g., Applegate 1986;
Chernoff \& Shapiro 1987) followed by Fokker--Planck (e.g., Chernoff
\& Weinberg 1990) and $N$-body simulations (e.g., Fukushige \& Heggie
1995; Portegies Zwart \etal\ 1998) have shown that mass loss due to
stellar evolution triggers an expansion, which can lead to rapid
cluster dissolution (see also Kalirai \& Richer 2010).

The extended surveys of Chernoff \& Shapiro (1987) and Chernoff \&
Weinberg (1990) showed that mass loss due to stellar evolution can
cause the rapid dissolution of clusters with a low initial
concentration and/or a flatter stellar initial mass function (IMF).

Fukushige \& Heggie (1995) used $N$-body simulations to explore the
early evolution of clusters dissolving due to mass loss associated
with stellar evolution. Their results are in general, qualitative
agreement with those of the Fokker--Planck models of Chernoff \&
Weinberg (see also Takahashi \& Portegies Zwart 2000 for an extensive
comparison between $N$-body simulations and Fokker--Planck models).
Fukushige \& Heggie also explored the mechanism behind this rapid
cluster dissolution and showed it to be the result of a loss of
equilibrium as the cluster expands and reach structural properties for
which no virial equilibrium is possible.

The expansion triggered by early mass loss due to stellar evolution is
stronger if the cluster is initially mass segregated. For a given
amount of mass loss, preferentially removing mass from the central
regions (where massive stars tend to be located in mass-segregated
clusters) increases the heating and strengthens the subsequent
expansion. $N$-body simulations by Vesperini \etal\ (2009) show that
as the degree of initial mass segregation increases, so does the
strength of the initial cluster expansion.  As a result, clusters
differing only in the degree of initial mass segregation can have very
different lifetimes.  Mackey \etal\ (2007, 2008) showed that the
stronger early expansion of mass-segregated clusters, along with the
subsequent heating from a population of stellar black holes, can
explain the radius--age trend observed for massive clusters in the
Magellanic Clouds.

A strong early loss of stars plays a key role in the evolution of
multiple stellar populations in clusters. Although different origins
have been suggested for the gas from which second-generation stars
might have formed (e.g., Decressin \etal\ 2001; Ventura \etal\ 2001;
see also Renzini 2008; and references therein), most models require
the first-generation population to have an initial mass which is at
least ten times higher than its current mass. As shown by D'Ercole
\etal\ (2008), early cluster expansion and mass loss can be
responsible for the escape of such a large fraction of
first-generation stars.

Finally, although this review focuses on the internal dynamics of
clusters rather than on cluster systems (see Harris 2010; Larsen 2010
in this volume), it is important to point out that evolution and
disruption of clusters due to gas expulsion and stellar evolution may
play an important role in the evolution of the mass function of
globular cluster systems (hereafter GCMF).

Young cluster systems have a power-law GCMF (e.g., Zhang \& Fall 1999)
while older clusters follow a bell-shaped distribution (e.g., Ashman
\& Zepf 1998; Harris 2001; Brodie \& Strader 2006). While evaporation
due to two-body relaxation can transform a power law into a
bell-shaped GCMF, most GCMF evolution models show that when GCMF
evolution is driven only by two-body relaxation, a galactocentric
dependence of the GCMF turnover not found in observational data is
also produced (e.g., Vesperini 1997, 1998, 2000, 2001; Baumgardt 1998;
Fall \& Zhang 2001; Vesperini \etal\ 2003; McLaughlin \& Fall 2008;
Baumgardt \etal\ 2008{\it a}; see Zepf 2008 for a recent review).

It has been shown that early cluster dissolution due to gas expulsion
and mass loss from stellar evolution can preferentially destroy
low-mass clusters and significantly flatten the low-mass end of the
initial power-law GCMF, without introducing any strong dependence on
galactocentric distance. Two-body relaxation and tidal shocks, while
still playing a key role and leading to disruption of a significant
number of clusters, will act on the GCMF already flattened by early
evolutionary processes without giving rise to any radial trend of GCMF
properties, inconsistent with observations (Vesperini \& Zepf 2003;
Baumgardt \etal\ 2008{\it a}; Parmentier \etal\ 2008).

\section{Long-term evolution of clusters}\label{sec:longterm}

For most old globular clusters, the core and half-mass two-body
relaxation timescales are shorter than the cluster ages. Two-body
relaxation plays a major role in driving the long-term evolution of
clusters and shaping their current structural properties, as well as
their stellar content.

In this section, we focus our attention on the effects of two-body
relaxation. We point out, however, that the evolution of clusters on
eccentric orbits and clusters in spiral galaxies can be further
affected by the tidal shocks during passages through the disc or near
the host galaxy's central bulge. Although in most cases two-body
relaxation is the dominant evolutionary process, depending on the
properties of the host galaxy and the cluster's orbital parameters,
tidal shocks can significantly speed up cluster dissolution and either
accelerate or slow down the evolution towards core collapse (e.g.,
Chernoff \etal\ 1986; Spitzer 1987; Weinberg 1994; Gnedin \etal\
1999). Encounters with giant molecular clouds and passages through
spiral arms can also affect cluster evolution and mass loss (Gieles
\etal\ 2006, 2007).

In the following subsections, we focus our attention on mass loss due
to two-body relaxation, the implications of mass loss for the
evolution of the cluster stellar mass function and the cluster's
structural evolution.

\subsection{Mass loss and cluster dissolution}

As a star exchanges energy with other single and binary stars in a
cluster due to close and distant encounters, it can reach an energy in
excess of the escape energy and escape from the cluster.

Mass loss due to two-body relaxation plays a key role in a broad range
of issues related to the evolution of globular clusters and globular
cluster systems. Specifically, for clusters surviving the early
evolutionary processes described in \S\ref{sec:early}, mass loss due
to two-body relaxation is the main process determining their
lifetimes. It affects the clusters' stellar content and stellar mass
function and determines the fraction of stars in the host galaxy's
field population contributed by clusters.

Although mass loss due to two-body relaxation has been the subject of
a large number of studies since the very early investigations of star
cluster dynamics (see, e.g., Heggie 2001 for a review, and references
therein), only recently a number of investigations have shed light on
some fundamental aspects of this process and its dependence on cluster
structural parameters and the strength of the host galaxy's tidal
field.

The external tidal field of the host galaxy plays a key role in the
process of mass loss. While isolated clusters undergo mass loss due to
the combined effects of close and distant encounters (e.g., Giersz \&
Heggie 1994; Heggie 2001; Baumgardt \etal\ 2002; Heggie \& Hut 2003),
their dissolution timescale is extremely long. The $N$-body
simulations of isolated star clusters of Baumgardt \etal\ (2002) show
that it takes about $10^3 t_\mathrm{rh}(0)$ for a cluster to lose
about half of its initial mass, where $t_\mathrm{rh}(0)$ is the
cluster's initial half-mass relaxation time.

For clusters evolving in the external tidal field of the host galaxy,
on the other hand, the mass-loss rate is much faster and the
dissolution time significantly shorter: by lowering the escape speed
and truncating the cluster sizes, the external tidal field
significantly enhances the mass-loss rate. The process of mass loss
due to two-body relaxation and the dissolution time associated with
this process depend on the strength of the external tidal
field. Fokker--Planck (e.g., Chernoff \& Weinberg 1990; Takahashi \&
Portegies Zwart 2000) and $N$-body simulations (e.g., Vesperini \&
Heggie 1997; Aarseth \& Heggie 1998) have shown that the mass-loss
rate and dissolution time, $T_\mathrm{d}$, of a cluster containing an
initial number of stars $N_\mathrm{i}$, evolving on a circular orbit
at a galactocentric distance $R_\mathrm{g}$ in a host galaxy with
circular velocity $v_\mathrm{c}$ is proportional to ${N_\mathrm{i}
R_\mathrm{g}\over \log(N_\mathrm{i}) v_\mathrm{c}}$. For given values
of $N_\mathrm{i}$, $v_\mathrm{c}$ and $R_\mathrm{g}$, the mass-loss
rate depends weakly on the cluster's internal structural properties
(e.g., concentration and half-mass radius).  An extensive survey of
$N$-body simulations was recently carried out by Gieles \& Baumgardt
(2008). Their results clearly show that, for a cluster with a given
$N_\mathrm{i}$, the strength of the tidal field is the dominant factor
determining a cluster's lifetime.

If the presence of the tidal field is properly modelled (as opposed to
being treated in a simplified way by introducing, for example, a
spatial or an energy cutoff), stars can escape the cluster only
through one of the Lagrangian points of the galaxy--cluster
system. Fukushige \& Heggie (2000) and Baumgardt (2001) showed that
the time needed for a star with an energy greater than the escape
energy to flow through one of the Lagrangian points and actually
escape from the cluster is not negligible and can, depending on the
star's orbital parameters, be even longer than a Hubble
time. Fukushige \& Heggie (2000), Baumgardt (2001) and Baumgardt \&
Makino (2003) showed that the escape timescale modifies the scaling of
the cluster dissolution time with cluster mass, and that
$T_\mathrm{d}$ is actually proportional to
$\left(N_\mathrm{i}/log(N_\mathrm{i})\right)^{0.75}$ (see also Lamers
\etal\ 2005 for analytical fits to the mass-loss rates resulting from
the $N$-body simulations of Baumgardt \& Makino 2003).

A number of studies based on $N$-body simulations have followed the
motion of escaping stars beyond the cluster's tidal radius and
explored the formation, structure and stellar content of the elongated
tidal tails that these stars populate and the relationship between
their orientation and the cluster's orbit (e.g., Combes \etal\ 1999;
Johnston \etal\ 1999; Dehnen \etal\ 2004; Fellhauer \etal\ 2007;
Montuori \etal\ 2007; Ernst \etal\ 2009; Odenkirchen \etal\ 2009).
Observational studies have detected the presence of tidal tails in
several clusters (e.g., Leon \etal\ 2000; Odenkirchen \etal\ 2003;
Grillmair \& Dionatos 2006) and also revealed the presence of density
clumps in the tidal tails (e.g., Leon \etal\ 2000). Similar clumps
were found by Capuzzo Dolcetta \etal\ (2005) in $N$-body simulations
of clusters on eccentric orbits. More recently, K\"upper \etal\ (2008)
and Just \etal\ (2009) carried out detailed studies of the orbits of
escaping stars and showed that overdensities along the tails are
expected even for clusters on circular orbits. The observed clumps can
be explained in terms of the epicyclic motion of escaping stars and a
slowdown in some portions of their orbits as they move away from the
cluster.

As we will discuss in the next subsection, many theoretical studies
have shown that mass loss due to two-body relaxation leads to the
preferential escape of low-mass stars. In agreement with these
results, Koch \etal\ (2004)---in an observational study of the Pal 5
main-sequence stellar luminosity function---found a significant
overabundance of low-mass stars in this cluster's extended tidal
tails.

\subsection{Mass loss and evolution of the stellar mass function}

As a result of the tendency towards energy equipartition, massive
stars tend to sink to the cluster's central regions, while low-mass
stars populate the outer parts and dominate the population of escaping
stars.

A consequence of the preferential loss of low-mass stars is that the
stellar mass function flattens as a cluster proceeds in its dynamical
evolution and loses mass. A number of $N$-body simulations have
revealed a close correlation between the fraction of the initial
cluster mass lost and the slope of the low-mass end of the stellar
mass function (Vesperini \& Heggie 1997; Baumgardt \& Makino 2003;
Trenti \etal\ 2009). Assuming that clusters are characterized by a
universal stellar IMF, the current slope of the mass function is a
good indicator of the extent to which a cluster has been affected by
dynamical evolution.

The relationship between a cluster's dynamical evolution, mass loss
and its stellar mass function is one of the manifestations of the
close interplay between different aspects of the astrophysics of star
clusters.  Observational investigations of the stellar mass function
of a number of Galactic globular clusters have found a range of
different slopes of the low-mass side of the present-day mass
function, which might indeed be the result of differences in the
clusters' dynamical histories (cf. Piotto \& Zoccali 1999).  A recent
observational study by De Marchi \etal\ (2007) resulted in a puzzling
trend between cluster structure and slope of the stellar mass
function: clusters with a flatter present-day mass function tend to
have low-concentration density profiles.  The origin of this trend is
not clear (see, e.g., Baumgardt \etal\ 2008{\it b} for a study
suggesting that primordial mass segregation might be required to
explain the extreme mass function flattening of some low-concentration
clusters) and will require further investigation of the evolution of
cluster structural properties as they evolve towards complete
dissolution.

A consequence of the preferential loss of low-mass stars is that the
evolution of a cluster's mass-to-light ratio, $M/L$, differs from that
driven solely by stellar evolution. Specifically, it has been shown by
Baumgardt \& Makino (2003; see also Lamers \etal\ 2006; Anders \etal\
2009 for recent studies of the photometric evolution of dissolving
clusters) that during most of a cluster's life, $M/L$ is lower than
what the cluster would have without any loss of stars.  Only at the
very end of its life, when the cluster approaches complete
dissolution, its stellar population becomes dominated by dark remnants
and $M/L$ rises to values greater than those expected if there were no
mass loss.  Kruijssen (2008) and Kruijssen \& Mieske (2009) further
investigated the effect of the preferential loss of low-mass stars on
$M/L$ and showed that the effects of mass loss can account for the
discrepancy between the observed $M/L$ values for a sample of Galactic
globular clusters and those expected on the basis of including only
the effects of stellar evolution.

Tidal shocks due to passages near the host galaxys's bulge for
clusters on eccentric orbits, and through the Galactic disc, will
inject energy into the cluster and further speed up the process of
mass loss due to two-body relaxation (e.g., Gnedin \& Ostriker 1997;
Gnedin \etal\ 1999). However, mass loss due to tidal shocks is
independent of stellar mass and only combined with mass segregation
can mass loss due to tidal shocks contribute to the flattening of the
stellar mass function (e.g., Vesperini \& Heggie 1997).

\begin{figure}[h]
\centering{
\includegraphics[width=8cm,angle=-90]{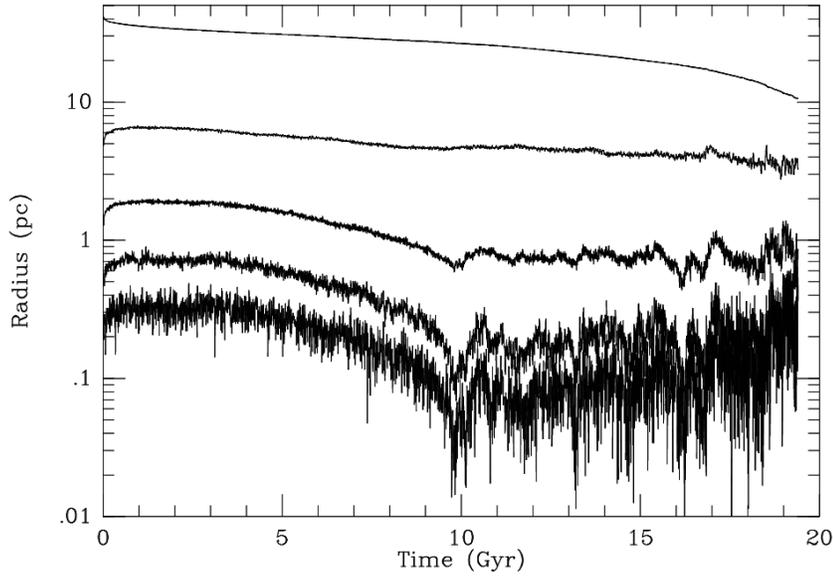}
}
\caption{Time evolution of the 0.1, 1, 10 and 50\% Lagrangian radii
  and tidal radius from an $N$-body simulation following the dynamical
  evolution of a star cluster orbiting at 4 kpc from the Galactic
  centre and with an initial mass of $M=1.49\times10^5$ M$_{\odot}$
  (from Aarseth \& Heggie 1998).  }
\label{fig2}
\end{figure}

\subsection{Structural evolution, core collapse and post-core collapse}

Figure~2 shows the results of an $N$-body simulation (from Aarseth \&
Heggie 1998) following the time evolution of the 0.1, 1, 10 and 50\%
Lagrangian radii and tidal radius of a star cluster orbiting at 4 kpc
from the Galactic centre and with an initial mass of $M_i\simeq
1.5\times 10^5$ M$_{\odot}$.  The initial expansion of the cluster is
a consequence of mass loss due to stellar evolution. As discussed in
\S\ref{sec:early}, clusters with low initial concentrations or a large
fraction of massive stars, or with strong initial mass segregation,
can undergo significant early expansion and rapidly dissolve as a
consequence of mass loss due to stellar evolution.

For clusters surviving early dissolution, figure 2 illustrates the
typical structural evolution towards core collapse driven by two-body
relaxation.  First studied by H\'enon (1961) using a Monte Carlo
integration of the Fokker--Planck equation, this process, also known
as gravothermal catastrophe, has subsequently been explored in a large
number of investigations (Antonov 1962 and Lynden--Bell \& Wood 1968
are among the first studies that explored the physics of this process;
see Heggie \& Hut 2003 for a review, and references therein). Without
intervention by an energy source balancing the loss of energy from the
core, this process would lead to a smaller and smaller core and a
diverging central density in a finite time.

Binary stars, either primordial or dynamically formed during close
encounters between single stars, can provide the energy needed to halt
core collapse and support the core in the post-core-collapse (PCC)
phase (see, e.g., Heggie \& Hut 2003; and references therein, also for
a review of the gravothermal oscillations that can characterize the
PCC phase of clusters).  In particular, as shown by Heggie (1975; see
also Heggie \& Hut 1993, 2003 for reviews and Goodwin 2010), binary
stars with binding energy, $\epsilon_\mathrm{b}$, moving in a stellar
system of single stars with mass $m$ and one-dimensional velocity
dispersion $\sigma$ such that $|\epsilon_\mathrm{b}|/m\sigma^2\gtorder
1$ will, on average, increase their binding energy and release the
energy lost to the star cluster.

The depth of core collapse and the concentration of a cluster in the
PCC phase, as measured, for example, by the ratio of the core to
half-mass radius, $r_\mathrm{c}/r_\mathrm{h}$, depends on the energy
source. Clusters supported by dynamically formed binaries will, in
general, be extremely concentrated, $r_\mathrm{c}/r_\mathrm{h} \approx
10^{-3}$ (e.g., Heggie \& Hut 2003).

The concentration of a cluster supported by primordial binaries
depends on the abundance of binaries and, to a smaller extent, on the
distribution of binary binding energy. Primordial binaries can support
a cluster in the PCC phase with values of $r_\mathrm{c}/r_\mathrm{h}$
as large as $\sim 0.05-0.08$ (see the analytical studies of Goodman \&
Hut 1989; Vesperini \& Chernoff 1994; and the Fokker--Planck and
$N$-body simulations of McMillan \etal\ 1990, 1991; Gao \etal\ 1991;
Heggie \& Aarseth 1992; Giersz \& Spurzem 2000, 2003; Heggie \etal\
2006; Fregeau \& Rasio 2007; Portegies Zwart \etal\ 2007; Trenti
\etal\ 2007{\it a}).

Observational studies have shown that about 20\% of Galactic clusters
have cuspy surface-brightness profiles and identified these objects as
clusters in the PCC phase (Djorgovski \& King 1986; Chernoff \&
Djorgoski 1989; Djorgovski \& Meylan 1994). As predicted by Chernoff
\& Shapiro (1987), PCC clusters are preferentially located near the
Galactic centre (Chernoff \& Djorgovski 1989). The smaller cluster
sizes set by the stronger tidal field at small galactocentric
distances lead to shorter relaxation times and more rapid evolution
towards core collapse.

While many clusters are still in the pre-core-collapse phase, a few
clusters have central and half-mass relaxation times which are short
enough that they should have already undergone core collapse (Chernoff
\& Djorgovski 1989; Trenti 2006).  These clusters, however, have
normal surface-brightness profiles, with a resolved core and values of
$r_\mathrm{c}/r_\mathrm{h}$ greater than can be supported even by a
large fraction of primordial binaries (and it is important to note
that recent observational studies suggest that the fraction of
primordial binaries in several clusters might only be $\approx 1-2$\%;
Davis \etal\ 2008{\it a}).

Alternative energy sources have been suggested.  Trenti (2006)
suggested that an IMBH at the centre of these clusters might support
the large values of $r_\mathrm{c}/r_\mathrm{h}$ (see Baumgardt \etal\
2005; Heggie \etal\ 2007; Trenti \etal\ 2007{\it b}; Gill \etal\ 2008;
Pasquato \etal\ 2009 for $N$-body simulations of clusters containing
an IMBH).  Heyl (2008), Heyl \& Penrice (2009) and Fregeau \etal\
(2009), following recent observational studies suggesting white dwarfs
might receive a kick velocity at the time of their formation (Davis
\etal\ 2008{\it b}; see also Kalirai \& Richer 2010), have explored
the effects of these kicks on cluster dynamics and showed that they
can indeed represent a significant energy source capable of delaying
core collapse and supporting high values of
$r_\mathrm{c}/r_\mathrm{h}$.

Hurley (2007) and Trenti \etal\ (2009) analysed the results of a
number of $N$-body simulations, following observational procedures,
and showed that adopting the observational definitions of clusters'
structural parameters reduces the differences between observational
and theoretical values of parameters such as the ratio of the core to
half-mass radius. As numerical models become more realistic, adopting
data-analysis procedures consistent with those used in observational
studies is important and will help to clarify for which clusters an
additional energy source is indeed required to explain the current
cluster structure.

The results of detailed $N$-body models for the dynamical evolution of
M4 and NGC 6397 (Heggie \& Giersz 2008, 2009{\it a,b}) show that there
are additional factors that can complicate the observational
identification of a cluster's dynamical phase. M4 and NGC 6397 are two
Galactic clusters with expected similar dynamical history which,
however, have different surface-brightness profiles: M4 has a normal
King profile (although of high concentration) while NGC 6397 exhibits
a cuspy profile and is usually classified as a PCC cluster. According
to the $N$-body models of Heggie \& Giersz (2008, 2009{\it a,b}), both
M4 and NGC 6397 are actually in the PCC phase and the differences in
the surface-brightness profiles are due to oscillations in the core
size which, in turn, cause the surface-brightness profile to oscillate
between cuspy and high-concentration King profiles. Depending on the
phase of this oscillatory behaviour, a cluster will be classified
either as a normal, high-concentration King model or a PCC cluster.

The results of all these studies suggest that it is likely that not
all clusters in the PCC phase have been classified as PCC
clusters. Further exploration of the range of possible observational
properties of PCC clusters and of additional observational indicators
of the PCC phase is necessary to ensure that the dynamical state of
normal and PCC clusters is correctly identified and that a proper
study of correlations between a cluster's dynamical state and other
internal (e.g., abundance of exotic objects such as pulsars, blue
stragglers, etc.) and external properties (e.g., position in the
Galaxy) can be carried out.

\section{Conclusions}\label{sec:concl}

The study of star cluster dynamics has a long history and significant
progress has been made in understanding the physics of the main
ingredients driving the dynamical evolution of star clusters.  Many
fundamental problems, however, are still open and new observations
continuously present us with new challenges and questions.

Understanding the role played by early evolutionary processes in the
dissolution of star clusters and in shaping the properties of those
which survive, how to identify a cluster's dynamical state, what
energy sources support clusters in the PCC phase and what the role is
of evolutionary processes in determining the current properties of
globular cluster systems are just a few of the issues that are still
awaiting firm solutions.

As the close interplay between dynamical evolution, the evolution of a
cluster's stellar content and the role played by the host galaxy's
environment has become clear, new and increasingly complex questions
crossing the boundaries of different fields have arisen.
Understanding the links between the formation and abundance of exotic
objects (such as blue stragglers, X-ray sources and IMBHs) and a
cluster's dynamical history, the relationship between a cluster's
stellar mass function, its structure and its past dynamical evolution,
the role played by the host galaxy's tidal field and the effect of its
time variation during galaxy assembly in shaping the current
properties of individual clusters and the global properties of cluster
populations are some examples of unsolved problems that require
increasingly sophisticated models. The observational evidence of the
presence of multiple stellar populations in globular clusters is the
latest major challenge and further illustrates the complexity of the
formation and dynamical evolution of star clusters.

As clusters evolve, their structural properties and stellar content
are modified by evolutionary processes. Clusters in different galaxies
and at different galactocentric distances have different dynamical
histories.  A better understanding of the effects of dynamical
evolution is an essential step in our attempts to shed light on the
relationship between current star cluster properties and those which
were imprinted by star-formation processes.

\begin{acknowledgements}
I acknowledge partial research support from NASA grant NNX08AH15G.
\end{acknowledgements}

\end{document}